\documentclass[showpacs,showkeys]{revtex4}
\usepackage{graphicx}

\usepackage{amsmath}
\usepackage{amssymb}
\usepackage{array}
\usepackage{color}

\begin{document}

\title{Asymptotically flat gravitating spinor field solutions. Step 2 - the compatibility of Dirac equations in a curve and a flat spaces}

\author{Vladimir Dzhunushaliev}
\affiliation{Institute for Basic Research,
Eurasian National University,
Astana, 010008, Kazakhstan \\
and \\
Institute of Physics of National Academy of Science
Kyrgyz Republic, 265 a, Chui Street, Bishkek, 720071,  Kyrgyz Republic}
\email[Email: ]{vdzhunus@krsu.edu.kg}

\date{\today}

\begin{abstract}
Using the fact that a spin connection is defined to an accuracy of a vector it is shown that the spin connection should be modified in such a manner that Dirac equation in a curve space would  be compatible with Dirac equation in a flat space.
\end{abstract}


\maketitle

\section{Dirac equation in a curve space}

This note is the continuation of the Ref.\cite{Dzhunushaliev:2009ny}. There it was mentioned that at the moment in general relativity  asymptotically flat solutions for a spinor field are unknown although do exist asymptotically flat solutions for all known fields: scalar and gauge fields. We continue the investigations in this direction and we will consider Dirac equation in a curve space. We will show that a spin connection should be modified by the addition of a vector in order to obtain correct Dirac equation in a flat space.

A covariant derivative of a spinor $\psi$ is
\begin{equation}
	\nabla_\mu \psi = \partial_\mu \psi - \Gamma_\mu \psi
\label{50}
\end{equation}
and the covariant derivative of a Dirac conjugated spinor $\bar \psi$ is
\begin{equation}
	\nabla_\mu \bar \psi = \partial_\mu \bar \psi + \bar \psi \Gamma_\mu
\label{55}
\end{equation}
where a spinor connection $\Gamma_\mu$ is defined from following equation (here we follow to Ref. \cite{poplawski}, section 1.7.2)
\begin{equation}
	\omega_{\bar a \bar b \mu}\gamma^{\bar a} \gamma^{\bar b} -
	\gamma_{\bar a} \Gamma_i \gamma^{\bar a} + 4\Gamma_\mu = 0.
\label{56}
\end{equation}
where $\gamma^{\bar a}$ are Dirac matrixes in a flat (Minkowski) space,
$\gamma^\mu = e_{\bar a}^{\phantom a \mu} \gamma^{\bar a}$ are Dirac matrixes in a curve space.
For the definition of tetrad, spin connection and so on we follow to Ref. \cite{poplawski}. The inverse tetrad $e_{\bar a}^{\phantom a \mu}$ satisfies
\begin{eqnarray}
	e^{\bar a}_{\phantom a \mu} e_{\bar b}^{\phantom b \mu} &=& \delta^{\bar a}_{\bar b},
\label{20}	\\
	e^{\bar a}_{\phantom a \mu} e_{\bar a}^{\phantom a \nu} &=& \delta^\nu_\mu .
\label{30}	
\end{eqnarray}
where ${\bar a}, {\bar b}=0,1,2,3$ are Lorentz indices; $\mu , \nu$ are world indices. The metric tensor $g_{\mu \nu}$ in a curve space is related to the Minkowski metric $\eta_{\bar a \bar b}$ through the tetrad
\begin{equation}
	g_{\mu \nu} = e^{\bar a}_{\phantom a \mu} e^{\bar b}_{\phantom a \mu} \eta_{\bar a \bar b}.
\label{40}
\end{equation}
The solution of \eqref{56} is
\begin{equation}
	\Gamma_\mu = - \frac{1}{4}\omega_{\bar a \bar b \mu}
	\gamma^{\bar a} \gamma^{\bar b} - A_\mu,
\label{57}
\end{equation}
where $A_\mu$ is a spinor-tensor quantity with one vector index. Substituting \eqref{57} to \eqref{56} gives us the equation for $A_\mu$
\begin{equation}
	-\gamma_{\bar a} A_\mu \gamma^{\bar a} + 4A_\mu = 0.
\label{60}
\end{equation}
The simplest solution of this equation is: $A_\mu$ is an arbitrary vector multiple of $I$. Usually $A_\mu = 0$. \textcolor{blue}{\emph{Here we would like to show that $A_\mu$ should be nonzero that the Dirac equation in a curve space would be consistent with the Dirac equation in a flat space.}} In order to find such solution we rewrite \eqref{57} in following form
\begin{equation}
	\Gamma_{\bar c} = e^{\phantom{\bar c} \mu}_{\bar c} \Gamma_\mu =
	- \frac{1}{4} \sum\limits_{\bar a, \bar b}
	\omega_{\bar a \bar b \bar c}	\gamma^{\bar a} \gamma^{\bar b} - A_{\bar c} =
	- \frac{1}{4}
	\sum_{\bar a, \bar b \neq \bar c} \omega_{\bar a \bar b \bar c}	
	\gamma^{\bar a} \gamma^{\bar b} -
	\frac{1}{2} \sum_{\bar a \neq \bar c} \omega_{\bar a \bar c \bar c}	
	\gamma^{\bar a} \gamma^{\bar c} - A_{\bar c}
\label{70}
\end{equation}
here there is not summation over $\bar c$. In order to find $A_{\bar c}$ we calculate the term $\gamma^{\bar c} \Gamma_{\bar c}$
\begin{equation}
	\gamma^{\bar c} \Gamma_{\bar c} = - \frac{1}{4}
	\sum_{\substack{c \\ a,b \neq c}} \omega_{\bar a \bar b \bar c}
	\gamma^{\bar c} \gamma^{\bar a} \gamma^{\bar b} + \frac{1}{2}
	\sum_{\substack{c \\ a \neq c}}
	\omega_{\bar a \bar c}^{\phantom{\bar a \bar c} \bar c} \gamma^{\bar a} -
	\sum\limits_{\bar a} \gamma^{\bar a} A_{\bar a}
\label{80}
\end{equation}
here we calculated 
\begin{equation}
	\sum_{\substack{c \\ a \neq c}}
	\omega_{\bar a \bar c \bar c} 
	\gamma^{\bar c} \gamma^{\bar a} \gamma^{\bar c} = 
	- \sum_{\substack{c \\ a \neq c}}
	\omega_{\bar a \bar c}^{\phantom{\bar a \bar c} \bar c} \gamma^{\bar a}.
\label{85}
\end{equation}
We choose
\begin{equation}
	A_{\bar a} = \frac{1}{2} \sum_{\bar c \neq \bar a}
	\omega_{\bar a \bar c}^{\phantom{\bar a \bar c} \bar c}.
\label{90}
\end{equation}
It is easy to show that \eqref{90} is the solution of \eqref{60}. Let us rewrite the term
\begin{equation}
	\sum_{\substack{c \\ a,b \neq c}} \omega_{\bar a \bar b \bar c}
	\gamma^{\bar c} \gamma^{\bar a} \gamma^{\bar b} =
	\sum_{\substack{c \\ a,b \neq c}} \omega_{\bar a \bar b \bar c}
	\gamma^{\bar a} \gamma^{\bar b} \gamma^{\bar c} .
\label{100}
\end{equation}
Finally we choose the spin connection $\Gamma_{\mu}$ in the form
\begin{equation}
	\tilde{\Gamma}_\mu = - \frac{1}{4} e^{\bar c}_{\phantom{\bar c} \mu}
	\sum_{\bar a, \bar b}\omega_{\bar a \bar b \bar c}
	\gamma^{\bar a} \gamma^{\bar b} -
	\frac{1}{2} e^{\bar a}_{\phantom{\bar a} \mu} \sum_{\bar c \neq \bar a}
	\omega_{\bar a \bar c}^{\phantom{\bar a \bar c} \bar c}
\label{110}
\end{equation}
here $\tilde{\Gamma}_\mu$ means that we use the Fock - Ivanenko coefficients
$\omega_{\bar a \bar b \bar c}$ with $\bar a \neq \bar b, \bar b \neq \bar c, \bar a \neq \bar c$. Consequently one can write Dirac equation in a curve space in following form
\begin{equation}
	i \gamma^\mu \tilde{\nabla}_\mu \psi - m \psi =
	i \gamma^\mu \left(
		\partial_\mu - \tilde{\Gamma}_\mu
	\right)\psi -
	m \psi = 0
\label{120}
\end{equation}
or
\begin{equation}
	i \left( \tilde{\nabla}_\mu \bar \psi \right)
	\gamma^\mu - m \bar \psi = i \bar \psi \left(
		\overleftarrow \partial_\mu + \tilde{\Gamma}_\mu
	\right) \gamma^\mu -
	m \bar \psi = 0
\label{130}
\end{equation}
for the Dirac conjugated spinor $\bar \psi$, here
$\bar \psi \overleftarrow \partial_\mu = \partial_\mu \bar \psi$. Usually Dirac equation is written as
\begin{equation}
	i \gamma^\mu \nabla_\mu \psi =
	i \gamma^\mu \left(
		\partial_\mu \psi - \Gamma_\mu
	\right)\psi - m \psi =
	i \gamma^\mu \left(
		\partial_\mu +
		\frac{1}{4} \omega_{\bar a \bar b \mu} \gamma^{\mu} \gamma^{\bar a} \gamma^{\bar b }
	\right) \psi - m \psi = 0.
\label{140}
\end{equation}
In order to compare equations \eqref{120} and \eqref{140} we will write Dirac operators
$\mathcal D_1, \mathcal D_2$ for both equations \eqref{120} and \eqref{140}
\begin{eqnarray}
	\mathcal D_1 \psi &=& \left( i 
		\gamma^\mu \partial_\mu +
		\frac{i}{4} \sum_{\bar a, \bar b, \bar c}\omega_{\bar a \bar b \bar c}
		\gamma^{\bar c} \gamma^{\bar a} \gamma^{\bar b} + 
		\frac{i}{2} \sum_{\bar b \neq \bar c, c}
		\omega_{\bar c \bar b}^{\phantom{\bar b \bar b} \bar b} \gamma^{\bar c} - m
	\right) \psi ,
\label{150} \\
	\mathcal D_2 \psi &=& \left( i \gamma^\mu 
		\partial_\mu +
		\frac{i}{4} \sum_{\bar a, \bar b, \mu}
		\omega_{\bar a \bar b \mu} \gamma^{\mu} 
		\gamma^{\bar a} \gamma^{\bar b } - m 
	\right) \psi
\label{160}
\end{eqnarray}
in Minkowski space for the spherical coordinate system. For the calculations of
$\omega_{\bar a \bar b \mu}$ we use following definitions from \cite{poplawski} (section 1.5.4)
\begin{eqnarray}
	\Lambda^{\bar a}_{\phantom{a} \mu \nu} &=& \frac{1}{2} \left(
		\partial_\nu e^{\bar a}_{\phantom a \mu} -
		\partial_\mu e^{\bar a}_{\phantom a \nu}
	\right) = - \Lambda^{\bar a}_{\phantom{a} \nu \mu},
\label{170}	\\
	\omega_{\alpha \beta \gamma} &=& - \Lambda_{\alpha \beta \gamma} +
	\Lambda_{\gamma \alpha \beta} - \Lambda_{\beta \gamma \alpha} = -
	\omega_{ \beta\alpha \gamma} ,
\label{180}	\\
	\omega_{\bar a \bar b \mu} &=& e_{\bar a}^{\phantom{a} \alpha}
	e_{\bar b}^{\phantom{b} \beta} \omega_{\alpha \beta \mu} =
	- \omega_{\bar b \bar a \mu}.
\label{190}	
\end{eqnarray}
The spinor ansatz we take in the simplest form
\begin{equation}
	\psi = e^{i \Omega t} \left(		
	\begin{array}{l}
		f(r)														\\
		0																\\
		i g(r) \cos \theta							\\
		i g(r) \sin \theta e^{i \phi}
	\end{array}
	\right).
\label{200}
\end{equation}
The metric is
\begin{equation}
	ds^2 = dt^2 - dr^2 - r^2 \left(
		d \theta^2 + \sin^2 \theta d \varphi^2
	\right).
\label{210}
\end{equation}
The Dirac matrices for the spherical coordinate system \eqref{210} are 
\begin{eqnarray}
	\gamma^{\bar 0} &=& 
	\begin{pmatrix}
		1 & 0 & 0 & 0		\\
		0 & 1 & 0 & 0		\\
		0 & 0 & -1 & 0	\\
		0 & 0 & 0 & -1
	\end{pmatrix} ,
\label{212}\\
	\gamma^{\bar 1} &=& 
	\begin{pmatrix}
		0 & 0 & \cos \theta & \sin \theta e^{-i \varphi}		\\
		0 & 0 & \sin \theta e^{i\varphi} & - \cos \theta 		\\
		-\cos \theta & -\sin \theta e^{-i \varphi} & 0 & 0	\\
		-\sin \theta e^{i \varphi} & \cos \theta & 0 & 0
	\end{pmatrix} ,	
\label{214}\\
	\gamma^{\bar 2} &=& 
	\begin{pmatrix}
		0 & 0 & -\sin \theta & \cos \theta e^{-i \varphi}		\\
		0 & 0 & \cos \theta e^{i\varphi} & \sin \theta 		\\
		\sin \theta & -\cos \theta e^{-i \varphi} & 0 & 0	\\
		-\cos \theta e^{i \varphi} & - \sin \theta & 0 & 0
	\end{pmatrix} ,	
\label{216}\\
	\gamma^{\bar 3} &=& 
	\begin{pmatrix}
		0 & 0 & 0 & -i e^{-i \varphi }		\\
		0 & 0 & i e^{i \varphi } & 0		\\
		0 & i e^{-i \varphi } & 0 & 0	\\
		-i e^{i \varphi } & 0 & 0 & 0
	\end{pmatrix} .	
\label{218}
\end{eqnarray}
The Fock - Ivanenko coefficients \eqref{190} are
\begin{equation}
	\omega_{\bar 1 \bar 2 2} = 1, \quad
	\omega_{\bar 1 \bar 3 3} = \sin \theta, \quad
	\omega_{\bar 2 \bar 3 3} = \cos \theta.
\label{220}
\end{equation}
For this case
\begin{equation}
	\omega_{\bar a \bar b \bar c} = 0 , \quad \text{with }
	\bar a, \bar b \neq \bar c
\label{225}
\end{equation}
which we will use in \eqref{150}. For the first case \eqref{150} we have
\begin{equation}
	\mathcal D_1 \psi = e^{i \Omega t}	\left(		
	\begin{array}{l}
		g'(r) + \frac{2}{r} g(r) + (m + \Omega)f(r)	\\
		0																				\\
		i \left[
			-f' + (\Omega - m) g(r)
		\right]	 \cos \theta										\\
		i \left[
			-f' + (\Omega - m) g(r)
		\right] \sin \theta	 e^{i \phi}																	
	\end{array}
	\right)
\label{230}
\end{equation}
that is agreed with Dirac equation for an electron in hydrogen atom. But for the second case \eqref{160} we have
\begin{equation}
	\mathcal D_2 \psi = e^{i t \Omega } \left(		
	\begin{array}{l}
		- \left[
			g'(r) + \frac{3}{r} g(r) + (m + \Omega ) f(r)
		\right]												\\
		-e^{i \varphi} \frac{\cot \theta}{2 r} g(r)		    \\
		-i \left[
			f'(r) + \frac{f(r)}{r} + (m - \Omega ) g(r)
		\right]	 \cos \theta								\\
		i \left[ - f'(r) +
		\frac{-3+\cos (2 \theta )}{\cos \theta \sin \theta} \frac{f(r)}{4r} -
		 (m - \Omega ) g(r)
		\right]	\sin \theta e^{i \varphi}								
	\end{array}
	\right)
\label{240}
\end{equation}
that obviously is not the necessary equation.

\section{Energy - momentum tensor}

According to Ref. \cite{ortin} the energy - momentum tensor for the classical spinor field is \footnote{It is necessary to note that the definition of Ricci tensors in Ref's \cite{poplawski} and \cite{ortin} have the opposite sign.}
\begin{eqnarray}
	T_{\mu \nu} &=& - \frac{i}{2} \left[
		\bar \psi \gamma_{( \mu} \nabla_{\nu )} \psi -
		\nabla_{( \mu} \bar \psi \; \gamma_{\mu )} \psi
	\right] + g_{\mu \nu} \mathcal L_{\psi},
\label{2-10} \\
	\mathcal L_{\psi} &=& \frac{i}{2} \left(
		\bar \psi \gamma^{\mu} \nabla_{\mu} \psi -
		\nabla_{\mu} \bar \psi \; \gamma^{\mu} \psi
	\right) - m \bar \psi \psi
\label{2-20}
\end{eqnarray}
here $a_{( \mu} b_{\nu )} = (1/2) (a_{\mu} b_{\nu} + a_{\nu} b_{\mu})$ means the symmetrization. For our goal we change $\nabla \rightarrow \tilde \nabla$
\begin{eqnarray}
	\tilde T_{\mu \nu} &=& - \frac{i}{2} \left[
		\bar \psi \gamma_{( \mu} \tilde \nabla_{\nu )} \psi -
		\tilde \nabla_{( \mu} \bar \psi \; \gamma_{\mu )} \psi
	\right] + g_{\mu \nu} \tilde{\mathcal L}_{\psi},
\label{2-13} \\
	\tilde{\mathcal L}_{\psi} &=& \frac{i}{2} \left(
		\bar \psi \gamma^{\mu} \tilde \nabla_{\mu} \psi -
		\tilde \nabla_{\mu} \bar \psi \; \gamma^{\mu} \psi
	\right) - m \bar \psi \psi .
\label{2-23}
\end{eqnarray}
The calculation of the energy - momentum tensor $\tilde T_{\bar a \bar b}$ with the spinor ansatz \eqref{200} and modified spin connection \eqref{110} gives us
\begin{eqnarray}
	\tilde T_{\bar 0 \bar 0} &=& - \Omega  \left[ f(r)^2 + g(r)^2 \right],
\label{2-30} \\
	\tilde T_{\bar 1 \bar 1} &=& - \left[
		g(r) f'(r) - f(r) g'(r)
	\right],
\label{2-40} \\
	\tilde T_{\bar 1 \bar 4} &=& \tilde T_{\bar 4 \bar 1} =
	\left[
		\Omega f(r) - \frac{g(r)}{2r}
	\right] g(r)	\sin \theta ,
\label{2-50} \\
	\tilde T_{\bar 2 \bar 2} &=& \tilde T_{\bar 3 \bar 3} = \frac{f(r) g(r)}{r}.
\label{2-60}
\end{eqnarray}
The current
\begin{equation}
	J^\mu = \bar \psi \gamma^\mu \psi
\label{2-65}
\end{equation}
is
\begin{eqnarray}
	J^0 &=& f(r)^2 + g(r)^2,
\label{2-70} \\
	J^3 &=& \frac{2 f(r) g(r)}{r} .
\label{2-80}
\end{eqnarray}
From \eqref{2-30} and \eqref{2-70} we see that there is the energy and charge densities. From \eqref{2-80} we see that there is the current $J^3$ along $\varphi$ direction leading to
$\tilde T_{\bar 1 \bar 4}$ component of energy - momentum \eqref{2-50}. One can say that
$\tilde T_{\bar 1 \bar 4} \neq 0$ and $J^3 \neq 0$ is the consequence of a spin distinct from zero.

\section{Conclusions}

Thus we have shown that the standard definition of the spin connection in a curve space should be modified that the Dirac equation in a curve space would have correct limit in going from a curve space to a flat space. The modification of the spin connection to lie in the fact that to the standard spin connection is a vector added. So that the modified spin connection has Fock - Ivanenko coefficents with unequal indices only.

\section*{Acknowledgements}

I am grateful to the Research Group Linkage Programme of the Alexander von Humboldt Foundation for the support of this research.

\end{document}